\documentclass[isre,nonblindrev]{informs3} 

\DoubleSpacedXI

\usepackage{natbib}
 \bibpunct[, ]{(}{)}{,}{a}{}{,}%
 \def\bibfont{\small}%

\usepackage{float}
\usepackage{makecell}
\usepackage[para]{threeparttable}
\usepackage{verbatim}
\usepackage{setspace}
\usepackage{algorithm}
\usepackage{algpseudocode}

\usepackage[T1]{fontenc}

\usepackage[whole]{bxcjkjatype} 

\TheoremsNumberedThrough     
\ECRepeatTheorems

\EquationsNumberedThrough    

\MANUSCRIPTNO{ISRE-0000-0000.00}

\renewcommand{\theARTICLETOP}{}
\begin{document}



\RUNTITLE{Chat-based Search Engine}


\TITLE{Crafting Knowledge: Exploring the Creative Mechanisms of Chat-Based Search Engines}

\ARTICLEAUTHORS{%
\AUTHOR{Lijia Ma\thanks{These authors contributed equally to the manuscript and are listed alphabetically.}}
\AFF{Michael G. Foster School of Business, University of Washington, Seattle, WA 98195, \EMAIL{lijiam@uw.edu}}
\AUTHOR{Xingchen (Cedric) Xu\footnotemark[2]}
\AFF{Michael G. Foster School of Business, University of Washington, Seattle, WA 98195, \EMAIL{xcxu21@uw.edu}} 
\AUTHOR{Yong Tan\thanks{Corresponding author}}
\AFF{Michael G. Foster School of Business, University of Washington, Seattle, WA 98195, \EMAIL{ytan@uw.edu}}
} 

\ABSTRACT{%
In the domain of digital information dissemination, search engines act as pivotal conduits linking information seekers with providers. The advent of chat-based search engines utilizing Large Language Models (LLMs) and Retrieval Augmented Generation (RAG), exemplified by Bing Chat, marks an evolutionary leap in the search ecosystem. They demonstrate metacognitive abilities in interpreting web information and crafting responses with human-like understanding and creativity. Nonetheless, the intricate nature of LLMs renders their ``cognitive'' processes opaque, challenging even their designers' understanding. This research aims to dissect the mechanisms through which an LLM-powered chat-based search engine, specifically Bing Chat, selects information sources for its responses. To this end, an extensive dataset has been compiled through engagements with New Bing, documenting the websites it cites alongside those listed by the conventional search engine. Employing natural language processing (NLP) techniques, the research reveals that Bing Chat exhibits a preference for content that is not only readable and formally structured, but also demonstrates lower perplexity levels, indicating a unique inclination towards text that is predictable by the underlying LLM. Further enriching our analysis, we procure an additional dataset through interactions with the GPT-4 based knowledge retrieval API, unveiling a congruent text preference between the RAG API and Bing Chat. This consensus suggests that these text preferences intrinsically emerge from the underlying language models, rather than being explicitly crafted by Bing Chat's developers. Moreover, our investigation documents a greater similarity among websites cited by RAG technologies compared to those ranked highest by conventional search engines. Collectively, these findings illuminate the unique sources, preferences, and potential economic implications of chat-based search engines' content selection behavior, providing fresh perspectives for both academic research and practical applications within the search engine domain.
}%

\KEYWORDS{Generative AI, RAG, AI Creativity, AI Metacognition, AI Alignment, Search Engine, NLP}

\maketitle
\newpage

``The medium is the message.'' - \cite{mcluhan1964understanding}

\section{Introduction}

In the contemporary digital landscape, the volume and variety of information are expanding rapidly. To bridge the gap between information providers and seekers, search engines have become indispensable tools in everyday life. Data from Serpwatch reveals that 87\% of mobile device users engage with search engines at least once daily. Google, for instance, processes an estimated 63,000 search queries per second, translating to approximately 2 trillion global searches each year\footnote{See \url{https://serpwatch.io/blog/search-engine-statistics/}.}. Given the dominant role of search engines in facilitating information access, website owners aspire to gain increased visibility on these platforms, whether for the consumption of their textual content or the purchase of products offered on their websites \citep{berman2013role, jansen2008determining}.

However, the search engine landscape is currently experiencing a paradigmatic shift owing to the emergence of Large Language Models (LLMs) and associated applications, especially those based on Retrieval Augmented Generation (RAG). On November 30, 2022, OpenAI introduced ChatGPT, a tool designed to address user queries in a conversational manner. Subsequently, Microsoft incorporated GPT-4 into its search engine, culminating in the debut of New Bing on February 7, 2023\footnote{See \url{https://techcrunch.com/2023/03/14/microsofts-new-bing-was-using-gpt-4-all-along/}.}.
This iteration of Bing integrates a chatbot, hereafter referred to as Bing Chat\footnote{It is now part of Microsoft Copilot. See \url{https://copilot.microsoft.com/}}, enabling online searches and the delivery of responses in comprehensive natural language. Concurrently, on March 21, 2023, Google commenced experimental trials of its chatbot-enhanced search engine, Bard, which has seen its monthly user base surpass 142 million as of early 2024\footnote{See \url{https://www.demandsage.com/google-bard-statistics/}.}. This novel category of chatbots, exemplified by Bing Chat and Bard, merges search engine capabilities with LLM-based generative artificial intelligence (AI), and are hence termed ``chat-based search engines''. In contrast, their predecessors are hereafter denoted as ``conventional search engines''.

Unlike conventional search engines that merely present a list of websites, chat-based search engines initiate their process by comprehending the query, subsequently assimilating information from pertinent websites, and ultimately delivering a coherent response that integrates references to multiple websites for substantiation, during which they demonstrate human-like metacognitive skills, including the acquisition of new knowledge and the demonstration of creativity. Nevertheless, the ``cognitive'' procedure through which the chatbot discerns pertinent information and formulates the final responses remains largely inscrutable, attributed to the complexity of its foundational model. As the parameter size of LLMs increases, elucidating the mechanism becomes increasingly unfeasible, even for the system's architects \citep{kosinski2023theory}. Delving into this machine-driven creative process is not only of scientific intrigue but also bears significant economic implications. Prior research in the realm of search engines has shed light on the considerable impact of search engine visibility on both the decision-making processes of individuals \citep{gong2018examining, ghose2014examining} and the distribution of welfare among stakeholders within an economic equilibrium \citep{berman2013role}. Within this novel search paradigm, the visibility of a website hinges on its citation by the chatbot. As chat-based search engines evolve and enhance their ability to modulate users' exposure to diverse websites\footnote{See \url{https://www.demandsage.com/google-gemini-statistics/}.}, the economic ramifications can escalate correspondingly.

Historically, conventional search engines also intentionally introduce opacity to their ranking algorithms, serving as a deterrent against manipulation by website proprietors. This obscurity, juxtaposed with the economically pivotal role of visibility through search engines, has given rise to the domain of Search Engine Optimization (SEO). SEO endeavors to decode the myriad factors that influence website standings within search engine results, aiming to tailor content to enhance its visibility \citep{reisenbichler2022frontiers, danaher2006factors}. Nonetheless, these insights pertaining to the ranking criteria of conventional search engines may not be directly applicable to comprehending the citation criteria of chat-based search engines, given their distinct attributes. To begin with, leading LLMs, such as those within the GPT series, are constructed using a human-in-the-loop methodology, wherein human evaluators determine the optimal responses to specific queries \citep{ouyang2022training}. Consequently, these LLMs tend to produce answers that conform to certain linguistic styles \citep{guo2023close}, potentially influencing their source preferences during information retrieval tasks. Furthermore, when referencing websites, LLMs must choose suitable textual content. However, LLMs demonstrate varied levels of understanding for different text types, typically evaluated through perplexity metrics \citep{gonen2022demystifying}. This disparity in comprehension may also shape their preferences in the source selection process. Given such inherent uncertainties, we posit our first research question.

\textit{RQ1: How do chat-based search engines select websites to cite in response to a query?} 

In order to address our first research question, we procure an extensive dataset by executing queries on New Bing and harvesting responses via Bing Chat (the chat-based interface), alongside the websites referenced therein and those listed by Bing (its conventional counterpart). Following a sequence of preprocessing and text mining procedures, we conduct a comparative analysis of the linguistic characteristics of websites cited within Bing Chat's responses against those that are not cited. Our findings indicate that the chat-based search engine exhibits a preference for source content that is more readable and analytical, while demonstrating lower levels of polarity and a reduced conversational tone. Furthermore, our analysis reveals a noteworthy trend: content deemed more ``readable'' by GPT, as quantified by perplexity scores, is more likely to be cited in support of the responses. This suggests the possibility that the criteria for citation are influenced by the characteristics of the underlying LLM, rather than being exclusively determined by the New Bing engineers.

Nonetheless, it is conceivable that these criteria may be derived from the conventional search engine engine, especially considering that Bing Chat initiates its process by first identifying relevant websites prior to learning and generating responses\footnote{See \url{https://www.microsoft.com/en-us/bing/do-more-with-ai/how-bing-chat-works}.}. Prior investigations into conventional search engines have unveiled distinct linguistic predilections within their ranking algorithms \citep{reisenbichler2022frontiers}. Should Bing Chat predominantly extract information from websites that are highly ranked by traditional search engines, it is likely that the ranking algorithms' preferences will be perpetuated, thus confounding the distinct preferences attributable to the foundational LLM. To discount this alternative explanation and gain a comprehensive understanding of how chat-based search engines diverge from traditional ranking algorithms, we propose our second research question.

\textit{RQ2: How do the selection criteria employed by chat-based search engines diverge from the ranking criteria utilized by conventional search engines?}

Utilizing the same dataset, we investigate the ranking criteria employing an identical set of variables. Our findings reveal that, while there are similarities between the ranking criteria of Bing Chat's citation process (such as readability), significant differences also exist in terms of other linguistic styles (e.g., certitude) and perplexity. These outcomes again underscore the distinct nature of chat-based search engines.

While our perplexity analysis hints at the influence of the foundational LLM on specific criteria, it remains uncertain whether these criteria are the product of deliberate design choices by Bing Chat's designers, akin to the dynamic anti-gaming measures employed in traditional ranking algorithms. However, compared with these traditional counterparts, the LLM used for chat-based search engines poses a greater challenge for manual design. Existing literature underscores that LLMs can manifest unforeseen behaviors once the scale of their parameters surpasses certain thresholds, with the underpinnings of these emergent phenomena eluding even the models' creators \citep{wei2022emergent,kosinski2023theory}. In light of these considerations, we advance our third research question.

\textit{RQ3: Do chat-based search engine's criteria originate through manual curation or naturally emerge from underlying language models?}

To address this research inquiry, we delve into the information retrieval and content creation processes executed by Bing Chat's foundational model, GPT-4, with a direct examination. Should the predilection for sourcing citation texts be attributed to the intrinsic model, analogous trends are expected to manifest in scenarios utilizing GPT-4 for Retrieval Augmented Generation (RAG). In pursuit of this objective, we gather an additional dataset through the use of OpenAI's knowledge retrieval API by prompting with both queries and a compilation of website contents. Upon replicating our examination of the citation criteria within this RAG context, we uncover that the criteria employed by the GPT-4 based RAG are qualitatively aligned with those observed in Bing Chat. This consistency further suggests that the citation preferences of Bing Chat are derived from the LLM's inherent capabilities, rather than being artificially engineered by the search engine's developers. Furthermore, the texts cited by RAG API demonstrate greater similarities when compared with the top-ranked websites presented by conventional search engines. The disparity provides another intriguing yet justifiable distinctiveness of the LLM-based information retrieval, given its responsibility for furnishing a singular coherent response, thereby requiring an elevated level of information congruity.

In the subsequent sections of this paper, we integrate relevant literature and highlight our theoretical contributions in Section 2. Section 3 describes our methodology for data acquisition and the construction of variables. Section 4 presents our data analysis, addressing each research question sequentially. Finally, Section 5 concludes with a discussion on practical implications and suggests directions for future research.

\section{Related Literature}
Our research builds on and contributes to three streams of literature: i) AI Performance, AI Metacognition, and AI Alignment; ii)search engine ranking, optimization, and marketing; iii) information retrieval with generative AI.

\subsection{AI Performance, AI Metacognition, and AI Alignment}
Broadly speaking, our paper is related to the research on AI performance, AI metacognition, and AI alignment, especially for LLM-based generative AI and related applications. 

Given the intricate architecture of LLM-based generative AI, the output and performance of such models is not entirely predictable by their designers. Particularly when the models scale up to a high number of parameters, they begin to exhibit emergent capabilities, which include unanticipated enhancements in performance on novel tasks, as discussed in the literature \citep{kosinski2023theory, johnston2023abstract, wei2022emergent, wei2022chain}. In response to these developments, computer scientists are devising novel metrics, tasks, and methodologies to comprehensively assess the efficacy of generative AI across various dimensions \citep{zhao2023survey}. For instance, \cite{wang2022super} compiles an extensive list of over 1,600 NLP tasks to evaluate the adaptability of universal language models.

More interestingly, recent advancements in LLMs have notably highlighted their potential in paving the way toward Artificial General Intelligence (AGI). According to a classification proposed by Google DeepMind, AI systems can be segmented into ten diverse categories, evaluated across five performance levels (emerging, competent, expert, virtuoso, or superhuman) and categorized by their breadth of application (narrow or general) \citep{morris2023levels}. Although earlier iterations of AI have evolved from rudimentary rule-based systems to highly advanced entities exemplified by AlphaFold \cite{jumper2021highly}, they are predominantly narrow in their focus and designed for specific tasks. In stark contrast, current generative AI models based on LLMs, like ChatGPT and Llama 2, demonstrate a broad spectrum of metacognitive capabilities, including the acquisition of new skills and creativity \citep{morris2023levels}. Motivated by these developments, scholars are now endeavoring to dissect the inherent cognitive processes of AI, such as personality and rationality \citep{mei2024turing, chen2023emergence, wang2023unraveling}, and to utilize AI's cognitive functions in simulating human responses \citep{brand2023using, horton2023large}.

As AI systems exhibit increasingly sophisticated cognitive capabilities and the capacity for creativity that surpasses human expectations, the concept of AI alignment has garnered heightened scrutiny \citep{gabriel2020artificial}. The objective of AI alignment is to ensure that AI systems act in accordance with human intentions and values, promoting robustness, interpretability, controllability, and ethicality in AI \citep{ji2023ai}. Throughout the development and application of Large Language Models (LLMs), alignment principles are integrally applied. For instance, during the development phase, Reinforcement Learning from Human Feedback (RLHF) is employed to align AI with human's goals \citep{ouyang2022training}. Subsequent to model development, AI outputs and their underlying mechanisms undergo evaluation and interpretation \citep{gilpin2018explaining}. Moreover, Red Teaming and other assurance strategies are deployed to enhance AI safety \citep{ganguli2022red}. Upon the deployment of AI systems, it is imperative from a societal alignment perspective to assess the systems' impacts on the broader society \citep{korinek2022aligned}. Within this expansive framework, previous research into the economics of AI can be subsumed under AI alignment, such as studies on the effects of Generative AI on labor markets \citep{liu2023generate, hui2023short}, social communities \citep{borwankar2023unraveling, xue2023can}, and software development \citep{peng2023impact, dohmke2023sea}.

Pursuant to this line of inquiry, our paper delves into the creative process underpinning the responses of chat-based search engines, assessing their outputs via an array of textual characteristics. Beyond identifying the distinct textual predilections of chat-based search engines, our study furnishes empirical support for the proposition that these preferences arise naturally from the foundational LLM, rather than being the product of manual curation. Furthermore, our findings elucidate potential mechanisms by which chat-based search engines may wield economic influence, specifically through their unique exposure allocation processes based on textual preferences, coupled with a reduction in information diversity.

\subsection{Search Engine Ranking, Optimization, and Marketing}

In terms of the application context, our inquiry also contributes to the extensive academic discourse on search engine rankings, optimization, and marketing. This body of work is multifaceted, encompassing a diverse array of subtopics.

Primarily, certain studies concentrate on the commercial values of search engine rankings and advertisements. These investigations endeavor to decode consumer reactions to search advertisements, examining phenomena such as rank-order effects, and to evaluate the efficacy of such advertisements \citep{ghose2014examining, goldfarb2011online, ghose2009empirical}. In addition, there is scholarly effort dedicated to elucidating the dynamics between various modes of information acquisition for consumers, such as organic versus sponsored searches \citep{yang2010analyzing}, generic versus branded searches \citep{rutz2011generic}, and the utilization of search engines in conjunction with social media platforms \citep{ghose2019modeling}.

Building upon these insights, the extant literature also delves into competitive strategies from the perspective of website proprietors (advertisers). It has been proposed that entities positioned at the pinnacle of organic search results may experience a reduction in their motivation to compete for sponsored clicks \citep{katona2010race}. Furthermore, modeling such competition among websites also necessitates the consideration of the actions of consumers and search engines, such as the search engine's incentive to prioritize low-quality websites to maximize its advertising revenue \citep{athey2011position, xu2011price, xu2012effects}. \cite{berman2013role} presents a framework that examines the impact of Search Engine Optimization (SEO) on the competitive landscape among advertisers and the resulting welfare for visitors, advertisers, and the search engine.

Finally, given the significance of search engine ranking, scholars also seek to employ techniques to comprehend and enhance website exposure. For organic search, previous research tends to investigate factors that can influence website ranking and subsequently optimize website content accordingly \citep{reisenbichler2022frontiers, erdmann2022search, ziakis2019important, su2014improve, danaher2006factors}. Regarding sponsored search, scholars concentrate on optimizing websites' keyword auction strategies to maximize revenue while adhering to specific constraints in a competitive environment \citep{liu2018semantic, du2017bidding, shin2015keyword, abhishek2013optimal, liu2010ex}.

Our investigation principally situates itself within the last stream of literature, as it too aims to elucidate the determinants influencing the content selection mechanism of search engines. Specifically, our study concentrates on the nascent domain of chat-based search engines, analyzing the criteria they employ to select websites for citation and how these criteria diverge from those of conventional search engines' ranking algorithms. This differentiation could engender a novel competitive environment owing to the unique traffic channel these chat-based search engines direct toward websites. Such an evolution also necessitates consideration in future SEO and platform strategies for both researchers and industry experts.
 
\subsection{Information Retrieval with Generative AI}

More specifically, our research further intersects with the rapidly expanding domain concerning information retrieval facilitated by generative AI. Despite the longstanding history of information retrieval methodologies \citep{schutze2008introduction, singhal2001modern}, the emergence of LLM-based generative AI and its associated applications are revolutionizing information retrieval processes. Consequently, scholars are endeavoring to design and evaluate the information retrieval techniques in this new context.

On the design side, the advent of LLM-based generative AI heralds a significant shift—or potentially a paradigmatic transformation—in the domain of information retrieval. These generative AI systems, trained on extensive datasets, are capable of delivering substantiated responses to queries, thereby facilitating direct access to information encapsulated within the original training corpus \citep{benedict2023gen}. Moreover, to embed domain-specific insights and contemporary information within these responses, designers may employ several strategies: (1) fine-tuning the LLMs with updated datasets \citep{hu2023llm}; (2) leveraging retrieval-augmented generation that combines the original model with new data \citep{hofstatter2023fid}; or (3) utilizing straightforward prompt engineering techniques with the incorporation of new data \citep{liu2023pre}. Significantly, LLM-based generative AI can be instrumental across various stages of information retrieval, including query reformulation, content recall, content re-ranking, and the synthesis of content for the final output \citep{zhu2023large}. This multifaceted utility paves the way for the enhancement or creation of novel applications powered by generative AI, such as LLM-based recommendation engines and search agents \citep{hua2023tutorial, nakano2021webgpt}.

Regarding the evaluation aspect, the intricacies inherent in both the foundational model and the unstructured information sources render the assessment of information retrieval techniques involving generative AI a formidable challenge. To address this, computer scientists devise an array of baseline tasks spanning multiple domains to gauge their efficiency \citep{chen2023benchmarking}. Additionally, human evaluations are conducted to appraise the systems' outputs \citep{lewis2020retrieval}. In certain instances, behavioral researchers engage human participants to undertake information-seeking tasks employing various information retrieval techniques, facilitating a comparative analysis of performance with human involvement \citep{xu2023chatgpt}.

Our research mainly enriches the second line of literature concentrating on the evaluation of LLM-based information retrieval systems. Employing a variety of NLP techniques and metrics, this study elucidates the selection criteria employed by Bing Chat for citing websites and delineates the divergence of these criteria from those utilized by traditional search engines' ranking algorithms. More crucially, our research reveals that the observed disparities stem from the foundational large language models, rather than being artifacts manually introduced by the designers of Bing Chat. Specifically, we observe that: (1) Bing Chat exhibits a preference for content that is more readily interpretable (i.e., perplexity) by models within the GPT family; (2) the application of retrieval-augmented generation with GPT-4 demonstrates qualitative consistency in selection criteria akin to that of Bing Chat. Consequently, our research also advocates for future scholarly endeavors to adopt a more integrative approach towards the evaluation of LLM-based applications and the underlying large language models.

\section{Data and Variables}
\label{sec:data}
\vspace{1em}
\subsection{Data Sources, Data Collection, and Preprocessing Procedures}

\subsubsection{Data Source 1: New Bing} \
\label{sec:raw_data_1}

We source the first raw dataset from New Bing, an AI-driven search engine developed by Microsoft, officially released on February 7, 2023. In contrast to traditional methods of listing pertinent web content, hereafter denoted as ``conventional search engine,'' New Bing offers users an additional chatbot known as ``Bing Chat'' which can provide consolidated responses by aggregating information from different websites\footnote{See \url{https://www.microsoft.com/en-us/edge/features/the-new-bing}.}. 

\begin{figure}[H]
\FIGURE{\includegraphics[width=6in]{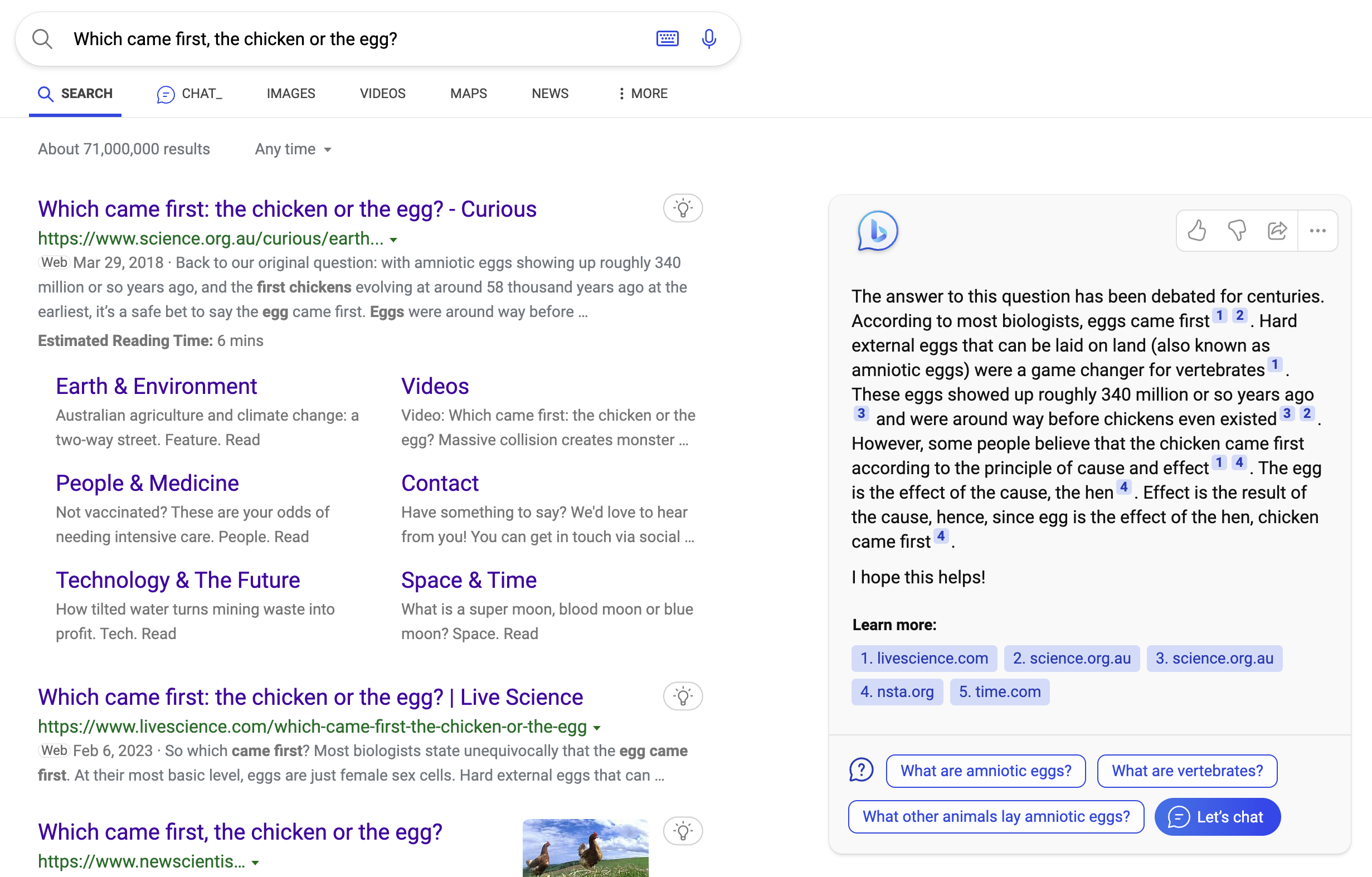}}
{An Example of Using New Bing \label{img:new_bing}}
{}
\end{figure}

Figure \ref{img:new_bing} offers a graphical depiction of the search output produced by the New Bing platform. To the left, conventional search outcomes are presented, each typically encompassing three elements: the website's title, its corresponding hyperlinks, and an excerpt from the site identified as most pertinent to the search query by the search algorithm. In juxtaposition, on the right-hand side, Bing Chat provides a composite response, integrating its pre-established knowledge database with contemporaneous data extracted from pertinent websites. Links to these sources are furnished at the dialogue box's base. Among these hyperlinks, some are explicitly referenced within the composite answer, whereas others are simply enumerated without direct citation. It is noteworthy that a single sentence may reference none to multiple websites, while a singular website may also be cited in support of none to multiple sentences.

To commence our analysis, we first extract and randomize all queries from the Human ChatGPT Comparison Corpus (HC3), a dataset introduced by \citet{guo2023close}, which is extensively utilized in subsequent Generative AI research for benchmarking and comparative studies \citep{zhao2023survey}. Each query is individually inputted into the search engine, acting as the keyword for information retrieval. A subset of queries fails to generate search results, attributable to the inclusion of sensitive or prohibited content, such as hate speech or violent material. Additionally, instances where the Bing Chat fails to produce responses due to the aforementioned reasons, or presents responses devoid of citations to websites, albeit infrequent, are also classified as unsuccessful queries. Constrained by the limitations of response time, our data collection ceases after conducting 700 successful queries that yield results. We then collect the conventional search outcomes displayed on the initial two pages of search results, encompassing titles, excerpts, and hyperlinks of the websites listed. Afterwards, for each Bing Chat response, we gather the textual content, the hyperlinks cited within these responses, and any specific locational markers referencing these hyperlinks. Furthermore, we proceed to capture and archive the complete textual content of all websites listed by the conventional search engine or cited by Bing Chat, culminating in a repository of 13,428 distinct websites. To offer a foundational understanding of the dataset's hierarchical structure, we present summary statistics delineated by various levels of granularity within the raw dataset in Table \ref{tab:summary_statistics_raw}.

For every query submitted, we show three variables that capture the scope of our data collection: $NumTotalWebs$, denoting the aggregate number of websites gathered per query; $NumCitedSentences$, signifying the count of sentences within each query's response that reference at least one website; and $NumCitedWebs$, specifying the number of websites cited by at least one sentence in Bing Chat's responses. The value of $NumCitedWebs$ naturally varies across different queries, reflecting Bing Chat's selection criteria to substantiate its responses. It is pertinent to highlight that the count of websites collected from the conventional search engine may also exhibit variability, attributed to the exclusion of duplicate websites that are displayed in the first two pages of conventional search outcomes.

Delving further, among the 700 responses generated by Bing Chat, a total of 2,586 sentences are identified that cite at least one website for reference. This observation is encapsulated in the variable $NumCitedWebsSent$, representing the count of websites cited at the sentence level. Correspondingly, for each website, we ascertain the frequency with which it is referenced across different sentences, thereby defining the variable $NumCitedSentencesWeb$ to quantify this metric.

\begin{table}[H] \centering 
\TABLE{Summary Statistics on Raw Data from Bing Chat\label{tab:summary_statistics_raw}}
{
\begin{tabular}{@{\extracolsep{10pt}}lrrrrrr} 
\\[-1.8ex]\hline 
\hline \\[-1.8ex] 
Statistic & \multicolumn{1}{c}{N} & \multicolumn{1}{c}{Mean} & \multicolumn{1}{c}{St. Dev.} & \multicolumn{1}{c}{Min} & \multicolumn{1}{c}{Median} & \multicolumn{1}{c}{Max} \\ 
\hline \\[-1.8ex] 
\textbf{Query-level Statistics}   \\
\textit{NumTotalWebs} & 700 & 19.18 & 1.18 & 12 & 20 & 20 \\ 
\textit{NumCitedSentences} & 700 & 3.69 & 1.73 & 1 & 3 & 14 \\ 
\textit{NumCitedWebs} & 700 & 3.37 & 1.31 & 1 & 3 & 10 \\
\textbf{Sentence-level Statistics}   \\
\textit{NumCitedWebsSent} & 2,586 & 1.38 & 0.67 & 1 & 1 & 5 \\ 
\textbf{Website-level Statistics}   \\
\textit{NumCitedSentencesWeb} & 13,428 & 0.27 & 0.68 & 0 & 0 & 8 \\ 
\hline \\[-1.8ex] 
\end{tabular} }
{}
\end{table}    


\begin{figure}[H]
\FIGURE{\includegraphics[width=5in]{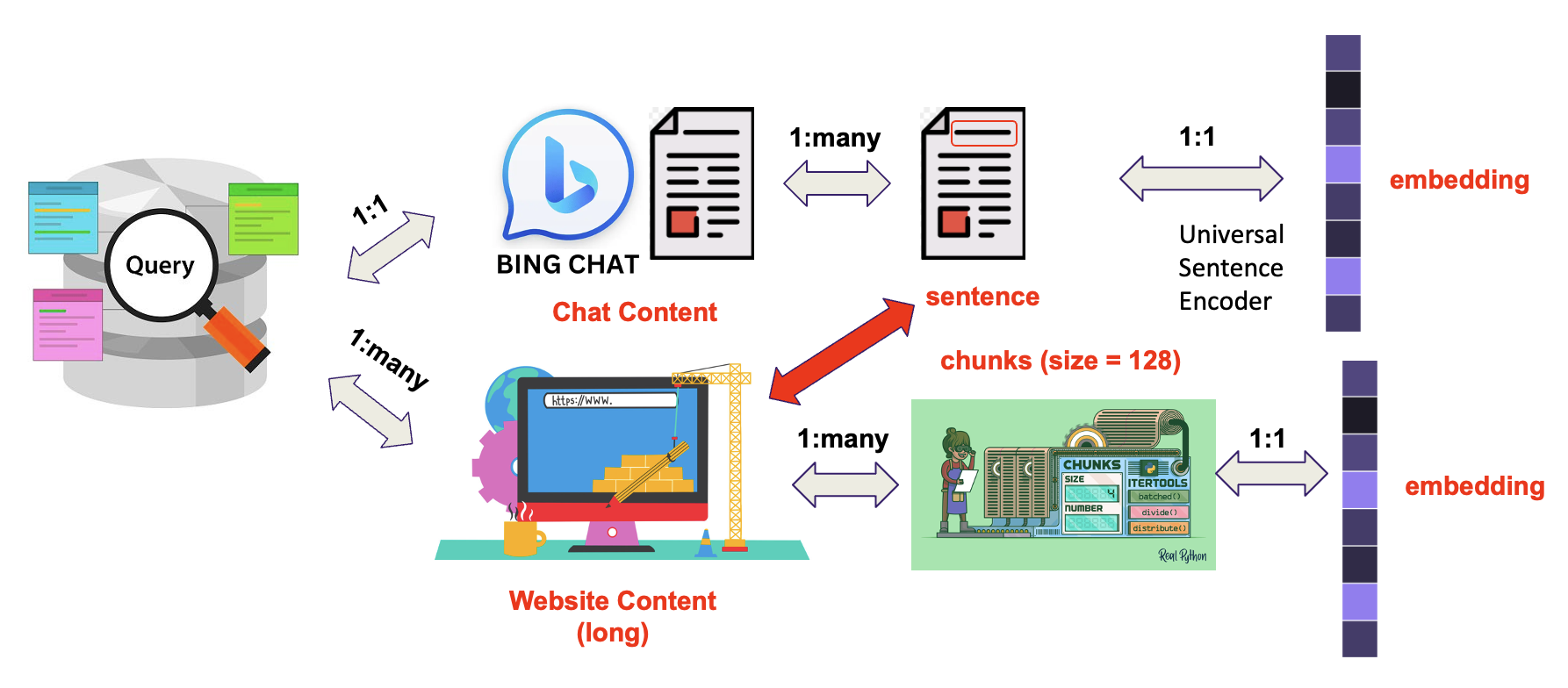}}
{Chunk Selection for Bing Chat Criteria Evaluation \label{img:preprocess}}
{}
\end{figure}

However, the dataset cannot be immediately employed for the investigation of Bing Chat's criteria, given that each website may comprise tens of thousands of words, with only a minor fraction being referenced to substantiate sentences in Bing Chat's responses. To find the most relevant texts, as delineated in the pseudocode Algorithm \ref{alg:chunk_1A} and Figure \ref{img:preprocess}, we undertake a series of preprocessing steps to compile dataset 1A: For every query, we pinpoint sentences within Bing Chat's response that cite at least one website. Subsequently, for each website sourced from this query, we identify the chunk exhibiting the highest similarity to each sentence with references and archive this chunk. This procedure involves dividing the websites into uniform chunks (128 characters each), employing the Universal Sentence Encoder to generate embeddings for both the sentences and all chunks from each website \citep{cer2018universal}, and utilizing cosine distance to gauge their similarities. This methodology yields a query-sentence-website level dataset, where each entry corresponds to a website chunk most akin to the sentence. For instance, should a query yield 20 websites and Bing Chat's response include 2 sentences that have citations, this results in 40 (20×2) distinct entries within dataset 1A. Using this method, we get 49,917 observations in total.

\begin{algorithm}[H]
\caption{Chunk Selection Process for Dataset 1A: Pseudocode}\label{alg:chunk_1A}
\begin{algorithmic}
\Require {Bing Chat's Responses, Website Content}

\State Initialize Dataset 1A.

\For{each query $i$:}

\State Retrieve all the sentences that cite at least one website $J_i$.

\State Retrieve all the related websites $K_i$.

\For{each sentence $j \in J_i$:}

\For{each website $k \in K_i$:}

\State Embed sentence $j$ using Universal Sentence Encoder.

\State Segment website $k$ into $M_k$ chunks of equal length.

\State Embed all the chunks $m \in M_k$.

\State Find the chunk $\hat{m} = \arg\max_{m \in M_k} \text{cosine\_similarity}(m'\text{'s embedding, } j\text{'s embedding)}$.

\State Add $i$, $j$, $k$, and $\hat{m}$'s content to Dataset 1A.

\EndFor
\EndFor
\EndFor
\end{algorithmic}
\end{algorithm}

Regarding our second research question, which seeks to elucidate the ranking criteria and compare them with Bing Chat's criteria, we assemble Dataset 1B from the websites cataloged by the conventional search engine. Given that the search engine highlights one excerpt for each listed website it deems most pertinent to the search query, we employ this excerpt as the basis for our ranking analysis. To extract the original content of these excerpts, we identify the chunk that includes the excerpt for each website listed by the conventional search engine, thereby creating Dataset 1B following the pseudocode Algorithm \ref{alg:chunk_1B}. This process results in 11,741 observations, each corresponding to one of the 11,741 websites. This quantity is less than the total number of websites gathered (i.e., 13,428), attributable to the fact that some websites are exclusively cited by Bing Chat and do not appear in the conventional search engine listings, thus precluding our ability to ascertain their ranking.

\begin{algorithm}[H]
\caption{Chunk Selection Process for Dataset 1B: Pseudocode}\label{alg:chunk_1B}
\begin{algorithmic}
\Require {Excerpts, Website Content}

\State Initialize Dataset 1B.

\For{each query $i$:}

\State Retrieve all the related websites $K_i$. 

\State Retrieve all the corresponding excerpts $E_k$ where $k \in CK_i$, where $CK_i \subseteq K_i$ contains websites that are listed in the conventional search engine results of query $i$. 

\For{each website $k \in K_i$:}

\State Segment website $k$ into $M_k$ chunks of equal length.

\State Find the chunk $\hat{m}$ that has the longest common subsequence with the excerpt.

\State Add $i$, $k$, and $\hat{m}$'s content to Dataset 2.

\EndFor
\EndFor
\end{algorithmic}
\end{algorithm}

\subsubsection{Data Source 2: GPT4-based RAG} \
\label{sec:raw_data_2}

To address our third research question, which endeavors to elucidate the origins of Bing Chat's preference for specific textual content -- questioning whether it is attributable to the inherent capabilities of the LLM or the result of deliberate modifications by the search engine's developers -- we propose a direct engagement with the foundational model, specifically GPT-4. OpenAI's Knowledge Retrieval API presents a viable means to accomplish this, offering the ability to systematically segment and access pertinent information from files provided by users, and to formulate responses that consider both the newly uploaded content and the pre-existing data repository\footnote{See \url{https://platform.openai.com/docs/assistants/tools/knowledge-retrieval}.}. By circumventing the bespoke pre-processing and post-processing layers implemented by Bing's designers for search engine tasks, this API may more authentically reveal the underlying LLM's linguistic predispositions.

To achieve our research objectives, it is imperative to construct an additional dataset, labeled as Dataset 2, building upon the queries and websites acquired from the first data source. For each query, we aim to upload the content of all the relevant websites and ascertain which of these are referenced by the aforementioned RAG API in formulating its responses. This approach, however, introduces a challenge analogous to the one encountered during the assembly of datasets 1A and 1B, namely, the prohibitive length and resultant computational cost of processing all the websites entirely. Consequently, a methodology to distill a more concise segment from each website is necessitated. Adhering to Algorithm 3, we divide each of the 13,428 websites into segments of uniform length (128 tokens in this instance). Subsequently, for each website indexed by the conventional search engines, we leverage the search result excerpt to identify and extract the chunk encompassing this excerpt as the representative chunk for the website\footnote{If a website is both listed by the conventional search engine and cited by Bing Chat, we also use this method.}. Conversely, if the website is exclusively cited by Bing Chat—thus lacking a corresponding excerpt—we utilize the sentence within Bing Chat's response that references the website as a focal sentence. The chunk bearing the highest similarity to this focal sentence, as quantified by cosine similarity between their embeddings, is then selected as the representative chunk for the website. Employing this methodology across the 13,428 websites, we successfully compile an equivalent number of corresponding chunks.

Subsequently, for each query, a compilation of websites is gathered alongside the query itself, with an equivalent number of chunks (one per website) extracted from these sites. These chunks are then amalgamated into a singular text document for each query, which functions as the retrieval source document. Upon submitting both the query content and the corresponding source text document to the OpenAI Knowledge Retrieval API, a synthesized response, inclusive of citations, is generated\footnote{A detailed illustration of the prompt, query, and resultant outputs is provided in Appendix \ref{appendix:rag}.}. Each citation delineates the original sentence along with its location within the submitted document, enabling the determination of whether a specific website (chunk) has been cited. Consequently, Dataset 2 comprises 13,428 entries, each accompanied with an outcome variable indicating its citation status by the RAG algorithm.

\begin{algorithm}[H]
\caption{Chunk Selection Process for Dataset 2: Pseudocode}\label{alg:chunk_2}
\begin{algorithmic}
\Require {Bing Chat's Responses, Excerpts, Website Content}

\State Initialize Dataset 2.

\For{each query $i$:}

\State Retrieve all the sentences that cite at least one website $J_i$.

\State Retrieve all the related websites $K_i$. 

\State Retrieve all the corresponding excerpts $E_k$ where $k \in CK_i$, where $CK_i \subseteq K_i$ contains websites that are listed in the conventional search engine results of query $i$. 

\For{each website $k \in K_i$:}

\State Segment website $k$ into $M_k$ chunks of equal length.

\If{website $k \in CK_i$}

    \State Find the chunk $\hat{m}$ that has the longest common subsequence with the excerpt.

\Else
    \State Find all the sentences from $J_i$ that cite website $k$. 

    \State Merge these sentences into a new sentence $j_{all}$
    
    \State Embed the sentence $j_{all}$ using Universal Sentence Encoder.

    \State Embed all the chunks $m \in M_k$.

    \State Find the chunk $\hat{m} = \arg\max_{m \in M_k} \text{cosine\_similarity}(m'\text{'s embedding, } j_{all}\text{'s embedding)}$.
    
\EndIf

\State Add $i$, $k$, and $\hat{m}$'s content to Dataset 2.

\EndFor
\EndFor
\end{algorithmic}
\end{algorithm}

\subsection{Variable Construction}

In the process of assembling the aforementioned three datasets (Datasets 1A, 1B, and 2) sourced from two distinct data repositories (New Bing and OpenAI's Knowledge Retrieval API), we systematically acquire the textual content alongside the corresponding outcome variables. This section elaborates on the variables derived from these datasets and presents summary statistics for each dataset.

We initiate by elucidating the outcome variables associated with each dataset. In Dataset 1A, for each sentence with references, we define the set of relevant chunks from all websites as the citation consideration set. Consequently, we obtain 49,917 observations, calculated as the product of the number of sentences with references and the number of websites retrieved for each query, summed across all queries, denoted as $\sum_i \text{Number of Sentences}_i \times \text{Number of Websites}_i$. In this dataset, the variable $Cited$ signifies whether a particular sentence cites the corresponding chunk, which, by extension, is equivalent to citing the website, given that a singular chunk from each website is considered. For Dataset 1B, we document 11,741 observations, where each encompasses the content of a website and its position as determined by a conventional search engine, encapsulated by the variable $Rank$. The content refers to the chunk within the website that contains the excerpt highlighted by the conventional search engine. In Dataset 2, consisting of 13,428 observations, each observation includes the content of a website and a variable $Cited$ denoting whether it is cited by the RAG API. Here, the content corresponds to a chunk which either contains the excerpt (if listed by the conventional search engine) or bears resemblance to sentences within Bing Chat's responses (if not listed).

In addressing the research questions pertinent to the citation and ranking behaviors of different systems, we also extract various features from website content (i.e., chunks) to serve as independent variables. Since our research is anchored in the domain of search engine ranking literature, prompting the inclusion of the $Readability$ metric, a widely recognized measure within this field \citep{reisenbichler2022frontiers}. $Readability$ evaluates the ease with which text can be understood by human readers. To quantify readability, we utilize the New Dale-Chall Readability Formula \citep{chall1995readability}, analyzing the readability of targeted website sections. For interpretative convenience, we introduce a negative sign to the formula's output, whereby a higher score implies enhanced readability.

Additionally, given the relevance of our research to assessing the outputs of generative AI, which aims to replicate human-like text, we examine several linguistic features commonly employed in such evaluations. The $Analytic$ variable quantifies the extent of analytical thought in the text, with higher values denoting greater logical consistency. The $Subjectivity$ measures the propensity to favor objective over subjective content, with higher scores reflecting greater subjectivity. These variables primarily concern the informational content's logicality and objectivity.

Moreover, we incorporate measures of expressive styles, such as $Polarity$, which gauges the emotional polarity (absolute value) of the text, with higher values indicating increased emotional polarization. The $Certitude$ variable assesses the text's level of assertiveness, for instance, texts employing phrases like "of course" may be deemed to exhibit higher certitude. The $Conversational$ measure evaluates the text's colloquialism, including the use of internet slang (e.g., ``lmao", ``lol") or discourse markers (e.g., ``um", ``oh"). The calculation of $Subjectivity$ and $Polarity$ is facilitated through TextBlob \citep{loria2018textblob}, while the remaining metrics are derived using the Linguistic Inquiry and Word Count (LIWC) tool \citep{boyd2022development}.

Finally, based on the research focus on LLM-based generative AI, we are particularly intrigued by the possibility that certain criteria might naturally emerge from these models. Of specific interest is the influence of ``readability" from the perspective of language models on the content selection process. To explore this, we introduce the concept of $Perplexity$, which is formally defined as the exponentiated average negative log-likelihood of a sequence as predicted by a language model \citep{azzopardi2003investigating}. Intuitively, it can measure the extent to which a model anticipates the input content, wherein a sentence that aligns closely with the model's expectations exhibits lower perplexity, and conversely, a more unexpected sentence exhibits higher perplexity.

Following the above steps, for every observation within the datasets, we derive a single outcome variable and seven independent variables, the summary statistics of which are presented subsequently. Dataset 1A, 1B, and 2 will be utilized for conducting regression analyses aimed at investigating research questions 1, 2, and 3, respectively.

\begin{table}[H] \centering 
\TABLE{Summary Statistics on Dataset 1A - Bing Chat's Citation\label{tab:summary_statistics_1a}}
{
\begin{tabular}{@{\extracolsep{10pt}}lrrrrrr} 
\\[-1.8ex]\hline 
\hline \\[-1.8ex] 
Statistic & \multicolumn{1}{c}{N} & \multicolumn{1}{c}{Mean} & \multicolumn{1}{c}{St. Dev.} & \multicolumn{1}{c}{Min} & \multicolumn{1}{c}{Median} & \multicolumn{1}{c}{Max} \\ 
\hline \\[-1.8ex] 
\textbf{Outcome Variable}   \\
\textit{Cited} & 49,917 & 0.07 & 0.26 & 0 & 0 & 1 \\ 
\textbf{Independent Variable}   \\
\textit{Readability} & 49,917 & -23.60 & 10.57 & -54.54 & -24.15 & 106.16 \\ 
\textit{Analytic} & 49,917 & 75.27 & 22.30 & 1 & 82.03 & 99 \\ 
\textit{Certitude} & 49,917 & 0.43 & 0.75 & 0 & 0 & 14.06 \\ 
\textit{Subjectivity} & 49,917 & 0.44 & 0.12 & 0 & 0.44 & 1 \\
\textit{Polarity} & 49,917 & 0.12 & 0.09 & 0 & 0.10 & 0.83 \\ 
\textit{Conversation} & 49,917 & 0.16 & 0.65 & 0 & 0 & 47.76 \\ 
\textit{Perplexity} & 49,917 & 4.72 & 0.35 & 1.96 & 4.74 & 6.74 \\
\hline \\[-1.8ex] 
\end{tabular} }
{}
\end{table}    

\begin{table}[H] \centering 
\TABLE{Summary Statistics on Dataset 1B - Conventional Bing's Ranking\label{tab:summary_statistics_1b}}
{
\begin{tabular}{@{\extracolsep{10pt}}lrrrrrr} 
\\[-1.8ex]\hline 
\hline \\[-1.8ex] 
Statistic & \multicolumn{1}{c}{N} & \multicolumn{1}{c}{Mean} & \multicolumn{1}{c}{St. Dev.} & \multicolumn{1}{c}{Min} & \multicolumn{1}{c}{Median} & \multicolumn{1}{c}{Max} \\ 
\hline \\[-1.8ex] 
\textbf{Outcome Variable}   \\
\textit{Rank} & 11,741 & 8.97 & 4.96 & 1 & 9 & 20 \\ 
\textbf{Independent Variable}   \\
\textit{Readability} & 11,741 & -23.79 & 10.27 & -54.54 & -24.39 & 76.30 \\ 
\textit{Analytic} & 11,741 & 76.17 & 21.99 & 1 & 83.13 & 99 \\ 
\textit{Certitude} & 11,741 & 0.43 & 0.76 & 0 & 0 & 7.75 \\ 
\textit{Subjectivity} & 11,741 & 0.44 & 0.12 & 0 & 0.44 & 0.92 \\
\textit{Polarity} & 11,741 & 0.12 & 0.09 & 0 & 0.10 & 0.68 \\ 
\textit{Conversation} & 11,741 & 0.16 & 0.52 & 0 & 0 & 10.08 \\ 
\textit{Perplexity} & 11,741 & 4.70 & 0.35 & 2.12 & 4.72 & 5.77 \\
\hline \\[-1.8ex] 
\end{tabular} }
{}
\end{table}    

\begin{table}[H] \centering 
\TABLE{Summary Statistics on Dataset 2 - RAG's Citation\label{tab:summary_statistics_2}}
{
\begin{tabular}{@{\extracolsep{10pt}}lrrrrrr} 
\\[-1.8ex]\hline 
\hline \\[-1.8ex] 
Statistic & \multicolumn{1}{c}{N} & \multicolumn{1}{c}{Mean} & \multicolumn{1}{c}{St. Dev.} & \multicolumn{1}{c}{Min} & \multicolumn{1}{c}{Median} & \multicolumn{1}{c}{Max} \\ 
\hline \\[-1.8ex] 
\textbf{Outcome Variable}   \\
\textit{Cited} & 13,428 & 0.19 & 0.39 & 0 & 0 & 1 \\ 
\textbf{Independent Variable}   \\
\textit{Readability} & 13,428 & -23.43 & 10.40 & -54.54 & -24.06 & 76.30 \\ 
\textit{Analytic} & 13,428 & 76.76 & 21.80 & 1 & 84.50 & 99 \\ 
\textit{Certitude} & 13,428 & 0.42 & 0.75 & 0 & 0 & 7.75 \\ 
\textit{Subjectivity} & 13,428 & 0.44 & 0.12 & 0 & 0.44 & 0.92 \\
\textit{Polarity} & 13,428 & 0.12 & 0.09 & 0 & 0.10 & 0.68 \\ 
\textit{Conversation} & 13,428 & 0.16 & 0.52 & 0 & 0 & 10.08 \\ 
\textit{Perplexity} & 13,428 & 4.70 & 0.35 & 2.12 & 4.72 & 6.74 \\
\hline \\[-1.8ex] 
\end{tabular} }
{}
\end{table}

\section{Data Analyses}
\label{sec:main}

In this section, we systematically address each of our research questions. First, in Section \ref{sec:bing_chat}, we scrutinize the criteria by which the chat-based search engine, specifically Bing Chat in our study, selects websites for inclusion in its responses. Following this, Section \ref{sec:conventional_rank} delineates the ranking criteria employed by conventional search engines and contrasts them with those utilized by Bing Chat. Finally, Section \ref{sec:RAG} elucidates the source of citation criteria by utilizing the RAG model based on GPT-4 and juxtaposing them with those employed by Bing Chat.

\subsection{Chat-based Search Engine's Content Selection Criteria}
\label{sec:bing_chat}

We initiate our exploration with the first research question, which investigates the criteria Bing Chat utilizes to select websites for substantiating its responses, employing Dataset 1A as elaborated in Section 3. For each query $i$, Bing Chat generates a response comprising at least one sentence that incorporates references. The crux of our analysis lies in discerning the textual predilections that predispose the selection of certain websites over others to underpin these referenced sentences. Given the voluminous content available on each website, we delineate a single relevant chunk from each website $k$ to constitute the consideration set for a given sentence $j$. Consequently, Dataset 1A forms a query-sentence-website level dataset, encompassing 49,917 entries.

The variable of interest, $Cited_{ijk}$, indicates whether a chunk from website $k$ has been referenced to support sentence $j$ in Bing Chat's response to query $i$. To decipher the relationship between the textual features of a website's chunk and its likelihood of citation, we incorporate a suite of independent variables, as delineated on the right side of Equation \ref{eq:1}. To strengthen the robustness of our results, we employ three distinct regression models: Ordinary Least Squares (OLS) regression (i.e., linear probability model), logistic regression, and probit regression. 

\begin{equation}
\label{eq:1}
\begin{split}
    Cited_{ijk} =& \beta_0 + \beta_1 * Readability_{ijk} + \beta_2 * Analytic_{ijk} + \beta_3 * Certitude_{ijk} + \\
    &\beta_4 * Subjectivity_{ijk} + \beta_5 * Polarity_{ijk} + \beta_6 * Conversation_{ijk} + \beta_7 * Perplexity_{ijk} + \epsilon_{ijk}
\end{split}
\end{equation}

\begin{table}[H]
\TABLE{Citation Criteria of Chat-based Search Engine \label{tab:bingchat}}
{
\begin{tabular}{@{\extracolsep{5pt}}lccc} 
\\[-1.8ex]\hline 
\hline \\[-1.8ex] 
 & \multicolumn{3}{c}{\textit{Dependent variable:}} \\ 
\cline{2-4} 
\\[-1.8ex] & \multicolumn{3}{c}{Cited} \\ 
\\[-1.8ex] & \textit{OLS} & \textit{Logistic} & \textit{Probit} \\ 
\\[-1.8ex] & (1) & (2) & (3)\\ 
\hline \\[-1.8ex] 
 Readability & 0.0009$^{***}$ & 0.0110$^{***}$ & 0.0057$^{***}$ \\ 
  & (0.0001) & (0.0016) & (0.0008) \\ 
  & & & \\ 
 Analytic & 0.0004$^{***}$ & 0.0073$^{***}$ & 0.0034$^{***}$ \\ 
  & (0.0001) & (0.0010) & (0.0005) \\ 
  & & & \\ 
 Certitude & $-$0.0014 & $-$0.0241 & $-$0.0105 \\ 
  & (0.0015) & (0.0260) & (0.0124) \\ 
  & & & \\ 
 Subjectivity & 0.0124 & 0.2084 & 0.0927 \\ 
  & (0.0109) & (0.1618) & (0.0790) \\ 
  & & & \\ 
 Polarity & $-$0.0240$^{*}$ & $-$0.3874$^{*}$ & $-$0.1689$^{*}$ \\ 
  & (0.0133) & (0.2118) & (0.1022) \\ 
  & & & \\ 
 Conversation & $-$0.0032$^{**}$ & $-$0.0637$^{**}$ & $-$0.0311$^{**}$ \\ 
  & (0.0014) & (0.0324) & (0.0154) \\ 
  & & & \\ 
 Perplexity & $-$0.0302$^{***}$ & $-$0.4185$^{***}$ & $-$0.2082$^{***}$ \\ 
  & (0.0036) & (0.0477) & (0.0240) \\ 
  & & & \\ 
 Constant & 0.2030$^{***}$ & $-$0.9414$^{***}$ & $-$0.6306$^{***}$ \\ 
  & (0.0184) & (0.2560) & (0.1254) \\ 
  & & & \\ 
\hline \\[-1.8ex] 
Observations & 49,917 & 49,917 & 49,917 \\ 
\hline 
\hline \\[-1.8ex] 
\end{tabular} 

}
{}
\begin{tablenotes}
\footnotesize
\centerline{Note: Robust standard errors in parentheses; *** $p$ $<$ 0.01, ** $p$ $<$ 0.05, * $p$ $<$ 0.1}
\end{tablenotes}
\end{table}

The regression outcomes, as presented in Table \ref{tab:bingchat}, manifest consistency across the three employed models, underscoring robust findings. First, congruent with prior literature on search engines \citep{reisenbichler2022frontiers}, chat-based search engines also exhibit a preference for content that is more readily comprehensible ($\beta_1 > 0$, $p < 0.01$). Additionally, Bing Chat demonstrate a proclivity for content characterized by higher degrees of logical reasoning and analytical thought ($\beta_2 > 0$, $p < 0.01$). Beyond these preferences more related to substantive information, Bing Chat also displays discernible inclination regarding stylistic attributes of content. Specifically, there is a tendency to reference websites manifesting reduced emotional polarity and a lower degree of conversationality ($\beta_5 < 0$, $\beta_6 < 0$), signaling a preference for more logical and formal expressions.

A particularly noteworthy finding pertains to Bing Chat's likelihood of citing websites with significantly lower perplexity scores ($\beta_7 < 0$, $p < 0.01$). Given that perplexity serves as a measure of text predictiveness from the perspective of the language model, this trend suggests an inclination towards sourcing content that resonates more closely with the foundational language model's training corpus. This also implies that the textual preferences exhibited by chat-based search engines may not be manually curated but rather emerge spontaneously from the underlying large language model, which is trained on extensive datasets.

\subsection{Comparison with Conventional Search Engines}
\label{sec:conventional_rank}

In the previous subsection, we delineate the selection criteria employed by Bing Chat for citing content in support of its responses. The objective of this subsection is to elucidate the ranking criteria utilized by conventional search engines, employing Dataset 1B for this purpose, and to undertake a comparative analysis with the criteria observed in Bing Chat. From a total collection of 13,428 websites, 11,741 are indexed by the conventional search engine and thus included in our analysis. This dataset also encompasses the websites that are both cited by Bing Chat and listed by the conventional search engine.

For each website $k$ listed by the conventional search engine in response to query $i$, we identify the chunk containing the excerpt highlighted by the engine for our ranking analysis. The dependent variable, $Rank_{ik}$, signifies the rank position of website $k$ within the conventional search engine results for query $i$. Notably, the ranking system commences at one, with lower values indicating superior rank positions. Following the same logic, we extract an identical set of independent variables for each website's chunk, as incorporated in the regression model delineated in Equation \ref{eq:2}. The ordered logit and ordered probit regression models are employed to ascertain the impact of these independent variables on the websites' ranking outcomes.

\begin{equation}
\label{eq:2}
\begin{split}
    Rank_{ik} =& \textit{Thresholds} + \beta_1 * Readability_{ik} + \beta_2 * Analytic_{ik} + \beta_3 * Certitude_{ik} + \\
    &\beta_4 * Subjectivity_{ik} + \beta_5 * Polarity_{ik} + \beta_6 * Conversation_{ik} + \beta_7 * Perplexity_{ik} + \epsilon_{ik}
\end{split}
\end{equation}

\begin{table}[H]
\TABLE{Ranking Criteria of Conventional Search Engine \label{tab:conv}}
{
\begin{tabular}{@{\extracolsep{5pt}}lccc} 
\\[-1.8ex]\hline 
\hline \\[-1.8ex] 
 & \multicolumn{3}{c}{\textit{Dependent variable:}} \\ 
\cline{2-4} 
\\[-1.8ex] & \multicolumn{3}{c}{Rank} \\ 
\\[-1.8ex] & \textit{Ordered Logit} & \textit{Ordered Probit} \\ 
\\[-1.8ex] & (1) & (2)\\ 
\hline \\[-1.8ex] 
 Readability & $-$0.0035$^{**}$ & $-$0.0018$^{**}$ \\ 
  & (0.0018) & (0.0010) \\ 
  & & & \\ 
 Analytic & 0.0004 & 0.0001 \\ 
  & (0.0008) & (0.0005) \\ 
  & & & \\ 
 Certitude & $-$0.0412$^{**}$ & $-$0.0247$^{**}$ \\ 
  & (0.0208) & (0.0124) \\ 
  & & & \\ 
 Subjectivity & 0.1841 & 0.1053 \\ 
  & (0.1504) & (0.0869) \\ 
  & & & \\ 
 Polarity & 0.1886 & 0.1619 \\ 
  & (0.1823) & (0.1062) \\ 
  & & & \\ 
 Conversation & 0.0690$^{**}$ & 0.0409$^{**}$ \\ 
  & (0.0280) & (0.0170) \\ 
  & & & \\ 
 Perplexity & $-$0.0512 & $-$0.0240 \\ 
  & (0.0457) & (0.0266) \\ 
  & & & \\ 
\hline \\[-1.8ex] 
Observations & 11,741 & 11,741 \\ 
\hline 
\hline \\[-1.8ex] 
\end{tabular} 

}
{}
\begin{tablenotes}
\footnotesize
\centerline{Note: Robust standard errors in parentheses; *** $p$ $<$ 0.01, ** $p$ $<$ 0.05, * $p$ $<$ 0.1}
\end{tablenotes}
\end{table}

Given that a higher rank value corresponds to a lower priority within the ranking system, it necessitates an inverse interpretation of the coefficients. As evidenced in Table \ref{tab:conv}, there are some observed similarities in preference between the conventional and chat-based search engines. Specifically, conventional search engines also favor websites characterized by greater readability ($\beta_1 < 0$, $p < 0.01$) and those exhibiting a less conversational tone ($\beta_6 > 0$, $p < 0.05$). However, contrary to Bing Chat, there is \textit{no} discernible preference for content imbued with analytical elements or content with diminished emotional polarity. Moreover, the conventional search engine demonstrates a preference for content with a higher degree of certitude ($\beta_3 < 0$, $p < 0.05$), a tendency \textit{not} mirrored by the chat-based counterpart. Notably, the analysis reveals \textit{no} significant correlation between perplexity and ranking, further highlighting the distinctiveness of LLM-based, chat-based search engines.

Therefore, the emergence of unique preferences within chat-based search engines underscores the imperative to revisit and reevaluate search engine research, particularly as this novel form of information retrieval continues to gain prominence.

\subsection{Criteria Foundation Exploration via Retrieval-Augmented Generation}
\label{sec:RAG}

In the preceding two subsections, we have identified distinctive selection criteria employed by the chat-based search engine, which notably diverges from the criteria of conventional search engine rankings. Intriguingly, Bing Chat's preference for content exhibiting lower perplexity intimates that such uniqueness may originate from the foundational large language model (specifically, GPT-4 in this study) rather than being deliberately programmed by New Bing's designers. Should this be the case, similar text selection criteria are anticipated to be observable during direct interactions with GPT-4 as well.

To further investigate this hypothesis and address our third research question, we compile Dataset 2 by utilizing OpenAI's knowledge retrieval API. This process entails the submission of queries alongside documents containing segments of content from various websites (the methodology for which is detailed in Section 3.1.2). The API's response, which includes references to specific segments of the document, enables us to ascertain whether the content from website $k$ is cited in response to query $i$. The outcome variable $Cite_{ik}$ is thus defined. Analogous to the approach outlined in the previous subsections, we employ the same spectrum of textual features extracted from the website chunks as independent variables. The analysis also encompasses three regression models (Linear probability, Logistic, and Probit regressions) with the specification summarized in Equation 3.

\begin{equation}
\label{eq:3}
\begin{split}
    Cited_{ik} =& \beta_0 + \beta_1 * Readability_{ik} + \beta_2 * Analytic_{ik} + \beta_3 * Certitude_{ik} + \\
    &\beta_4 * Subjectivity_{ik} + \beta_5 * Polarity_{ik} + \beta_6 * Conversation_{ik} + \beta_7 * Perplexity_{ik} + \epsilon_{ik}
\end{split}
\end{equation}

\begin{table}[H]
\TABLE{Citation Criteria of GPT-based RAG \label{tab:rag}}
{
\begin{tabular}{@{\extracolsep{5pt}}lccc} 
\\[-1.8ex]\hline 
\hline \\[-1.8ex] 
 & \multicolumn{3}{c}{\textit{Dependent variable:}} \\ 
\cline{2-4} 
\\[-1.8ex] & \multicolumn{3}{c}{Cited} \\ 
\\[-1.8ex] & \textit{OLS} & \textit{Logistic} & \textit{Probit} \\ 
\\[-1.8ex] & (1) & (2) & (3)\\ 
\hline \\[-1.8ex] 
 Readability & 0.0015$^{***}$ & 0.0096$^{***}$ & 0.0055$^{***}$ \\ 
  & (0.0004) & (0.0023) & (0.0013) \\ 
  & & & \\ 
 Analytic & 0.0004$^{**}$ & 0.0029$^{**}$ & 0.0016$^{**}$ \\ 
  & (0.0002) & (0.0012) & (0.0007) \\ 
  & & & \\ 
 Certitude & 0.0058 & 0.0386 & 0.0213 \\ 
  & (0.0047) & (0.0300) & (0.0171) \\ 
  & & & \\ 
 Subjectivity & 0.0391 & 0.2588 & 0.1413 \\ 
  & (0.0314) & (0.2030) & (0.1151) \\ 
  & & & \\ 
 Polarity & $-$0.0917$^{**}$ & $-$0.6135$^{**}$ & $-$0.3405$^{**}$ \\ 
  & (0.0373) & (0.2520) & (0.1432) \\ 
  & & & \\ 
 Conversation & $-$0.0121$^{*}$ & $-$0.0863$^{*}$ & $-$0.0478$^{*}$ \\ 
  & (0.0066) & (0.0498) & (0.0267) \\ 
  & & & \\ 
 Perplexity & $-$0.0323$^{***}$ & $-$0.2047$^{***}$ & $-$0.1179$^{***}$ \\ 
  & (0.0097) & (0.0606) & (0.0351) \\ 
  & & & \\ 
 Constant & 0.3408$^{***}$ & $-$0.5370$^{*}$ & $-$0.3424$^{*}$ \\ 
  & (0.0501) & (0.3152) & (0.1824) \\ 
  & & & \\ 
\hline \\[-1.8ex] 
Observations & 13,428 & 13,428 & 13,428 \\ 
\hline 
\hline \\[-1.8ex] 
\end{tabular} 
}
{}
\begin{tablenotes}
\footnotesize
\centerline{Note: Robust standard errors in parentheses; *** $p$ $<$ 0.01, ** $p$ $<$ 0.05, * $p$ $<$ 0.1}
\end{tablenotes}
\end{table}

Table \ref{tab:rag} elucidates that the RAG based on GPT-4 exhibits pronounced and consistent preferences towards text characteristics compared with Bing Chat. Notably, it demonstrates a higher likelihood of citing content that is more readable, logical, less polarized, less conversational, and more predictable by GPT models (evidenced by lower perplexity scores). These findings reinforce the argument that the criteria applied by chat-based search engines are derived from the intrinsic attributes of the underlying large language model, rather than being manually engineered. This underscores the importance of considering the foundational LLM in analyses of future chat-based search engine developments.

All preceding analyses have concentrated on textual characteristics, treating different websites as isolated entities. Nonetheless, the similarity of content across websites constitutes a significant dimension for evaluating search engine outputs. As underscored in previous studies \citep{reisenbichler2022frontiers}, content similarity can offer insights into the diversity of information made available to users and elucidate the exposure relationship among various websites. This metric assumes greater relevance in the context of chat-based search engines, which are tasked with providing a singular answer that references multiple supporting websites, thereby creating an indirect linkage among these sites through the response. Hence, in addition to website-level textual disparities, chat-based search engines may also exhibit unique ``market-level'' content similarities.

Employing Dataset 2, our investigation extends to the analysis of content similarity among websites cited through the RAG process in contrast to those ranked highly by traditional search engines. For each query \(i\) involving citations from \(N_i\) websites, we calculate the mean cosine similarity among the embeddings of these cited sites. In a parallel fashion, for the top \(N_i\) websites as per conventional search rankings, their average cosine similarities are computed. Out of 700 queries, 607 that feature at least two citations during the RAG process are selected for this similarity analysis, resulting in two sets of 607 similarity metrics—one derived from RAG-cited websites and the other from websites top-ranked by conventional search engines. A t-test between these datasets unveils a notable finding: websites cited by RAG display a higher degree of similarity compared to those in the top-ranked group by traditional search engines (\(t-stats = 5.32\), \(p < 0.01\)), with the former's mean similarity at 0.570 (\(std = 0.117\)) and the latter's at 0.533 (\(std = 0.128\)).

Thus, in addition to distinctive criteria based on textual features at the individual website level, applications through RAG also manifest a reduction in information diversity among cited websites relative to an equivalent number of top-ranked websites. These findings collectively underscore the unique operational mechanisms and outputs of chat-based search engines, underscoring the imperative for revised industry practices and further research inputs.

\section{Discussion and Conclusion}

With the advent of LLMs, a novel chat-based search engine paradigm has been introduced, leveraging LLMs to assimilate information from web sources and articulate responses in natural language, thereby exhibiting human-like cognitive faculties such as information assimilation and the generation of novel content. In pursuit of delineating the criteria for information selection and the foundational principles thereof, this study compiled two datasets through engagements with New Bing and the Knowledge Retrieval API powered by GPT-4. Our findings elucidate Bing Chat's inclination for content that is not only readable, formal, and logical but also exhibits lower perplexity, indicating a preference for text congruent with the anticipations of the underlying LLM. This preference is not adequately elucidated by traditional search engine ranking algorithms, underscoring the distinctive nature of chat-based search engine outputs. Additionally, through the employment of an ancillary dataset acquired via the RAG API, our research substantiates a consistent textual preference that aligns between the RAG API and Bing Chat, suggesting that these preferences are inherent to the language models rather than being explicitly configured by Bing Chat developers. Beyond the aforementioned website-level distinctions, our investigation discerns a greater similarity among websites referenced by RAG technologies vis-à-vis those prioritized by conventional search engines, signaling a broader market-level uniqueness.

Our research intersects with and contributes to multiple domains within the scholarly literature. Predominantly, our study falls into the AI alignment framework, which emphasizes the interpretation of AI systems' behavior. We enhance this body of work by assessing the distinctive outputs generated by a LLM-based AI system and delineating the mechanisms driving these outputs. More specifically, our investigation not only reveals how chat-based search engines reference websites based on textual characteristics, but also furnishs evidence that such selection criteria are influenced by the foundational LLM. This insight extends the literature on evaluating information retrieval systems powered by generative AI. Additionally, we underscore the importance of considering both LLM-based applications and the underlying LLMs in tandem to gain a comprehensive understanding of these systems' behaviors. Finally, our study further enriches search engine literature by delineating the criteria of a novel form of search engine, marking a preliminary exploration into SEO practices for chat-based search engine. We also demonstrate that the chat-based search engine's imperative to deliver coherent responses, as opposed to simply listing websites, reduces the information diversity among highly visible websites. This observation may inform future economic analyses within the search engine ecosystem.

Our study is not merely of scientific interest but also offers substantive insights for a diverse array of stakeholders. For developers of LLM-based RAG systems, including chat-based search engines, direct interaction with the underlying model is advised. This approach allows for a nuanced understanding of behaviors potentially attributed to the foundational LLM, enhancing system testing and refinement. Similarly, this holds significance for SEO professionals and website proprietors aiming to amplify their visibility within chat-based search engine results. By engaging directly with the LLM, they can fine-tune and evaluate their content prior to website updates. Furthermore, given the chat-based search engine's response format, it could be beneficial to optimize for compatibility with frequently cited websites. Users, on the other hand, should be cognizant of the limited information diversity inherent to chat-based search engine's responses, necessitating a balance between search efficiency and the breadth of information. Furthermore, the coexistence of chat-based and traditional search engines necessitates adaptive behavior from both website owners and users, given the distinct criteria of each search modality. This dynamic introduces additional market complexities, urging market administrators and regulatory bodies to re-evaluate market designs, such as auction mechanisms, to facilitate information dissemination and balance stakeholder welfare effectively.

\newpage
\bibliography{thebibliography}

\bibliographystyle{informs2014}

%
%
%





\newpage 

\begin{APPENDICES}

\section{A Sample of GPT4-based RAG Result}
\label{appendix:rag}
\subsection*{Sample Prompt} 
Please answer the question based on the content in the file attached: 
\{ query content \} Please answer the question with several clear references to the original files.

\subsection*{Sample Query} 
Why do we wear ties? Like why as in the origins, not just to look professional. Who said it made us look professional? 

\subsection*{Sample Answer from GPT-4 API} 
The origins of wearing ties trace back to various historical points and represent different meanings and purposes throughout history:

- **Ancient Roots in China**: Ties, in a form akin to scarves, were worn by ancient Chinese warriors; terracotta statues that are around 2000 years old display warriors with silk around their necks, indicating that ties were a sign of power and to establish identity even in ancient times【7†source】.

- **Croatian Mercenaries' Influence**: The tie gained popularity during the 30-year war in the 17th century when Croatian mercenaries who fought alongside King Louis XIII of France wore pieces of cloth around their neck to tie their jackets; King Louis XIII himself took a liking to this practice, which helped to popularize the accessory【8†source】.

- **Roman Era Representation**: Soldiers depicted on Trajan's Column, a monument commemorating Roman Emperor Trajan's victory over the Dacians, are shown wearing ties as a symbol of a warrior's courageous character, providing further historical evidence of ties being associated with military and strength【9†source】.

- **Elitism and Professionalism**: In Victorian England, guidelines on how to properly wear and tie a necktie emerged, such as the advice found in the book "Clothes And The Man," which could imply a sense of professionalism and elitism associated with the proper way to wear a tie【10†source】.

- **Designer Trends in the 1920s**: The modern notion of the tie as a professional and decorative accessory might be traced to the 1920s, where French fashion designers innovated the "designer tie" utilizing expensive fabrics and textile patterns to create stylish and coveted pieces【11†source】.

These references suggest that the wearing of ties has evolved from ancient symbols of power and identification to part of military uniform, and finally into a sophisticated fashion item that carries connotations of professionalism, status, and social identity. The tradition of wearing ties has been influenced by various cultures and periods, eventually leading to the contemporary association of ties with professional attire.

\subsection*{Sample Annotations}

【7†source】

\noindent \textbf{Quote}: Ancient Chinese warriors are the first known wearers of scarves. A discovery of 2000-year-old terracotta statues shows the warriors wearing silk around their necks. This shows that even back then, ties were a sign of power and to show identity. 

\noindent【8†source】

\noindent \textbf{Quote}: During the 30 year war, Croatian mercenaries fought alongside the King in battle. To keep their jackets tied they used a piece of cloth which King Louis XIII took a strong liking to ...

\noindent【9†source】

\noindent \textbf{Quote}: When Roman emperor Trajan defeated the Dacians, he erected a marble column to celebrate his victory. This column featured thousands of soldiers, many of whom were wearing ties, again as a symbol of a warrior’s courageous character. 

\noindent【10†source】

\noindent \textbf{Quote}: When it came to bow ties, the major was very clear: ``Of course, no gentleman ever does wear a made-up tie.''

\noindent【11†source】

\noindent \textbf{Quote}: In the 1920s, French fashion designers innovated the ``designer tie,'' which was made from expensive fabrics and decorated with textile patterns.

\subsection*{Interpretation of the Results}
In this interaction, we submit a query along with a document comprising chunks extracted from all the corresponding websites. The API subsequently furnishes a cohesive response, supported by four distinct references, each delineating the original portions (i.e., annotations) sourced from the uploaded document. Subsequently, we leverage these annotations to identify the websites containing these four annotations, designating them as cited websites, while categorizing the remaining websites as uncited.

\end{APPENDICES}

\end{document}